\documentstyle[aps,epsf]{revtex}  

\begin{document}        

\baselineskip 14pt

\newcommand{\gam}{$\gamma$}
\newcommand{\Cv}{Cherenkov\ }
\def\lsim{\mathrel{\hbox{\rlap{\hbox{\lower4pt\hbox{$\sim$}}}\hbox{$<$}}}}
\def\gsim{\mathrel{\hbox{\rlap{\hbox{\lower4pt\hbox{$\sim$}}}\hbox{$>$}}}}

\title{High Energy Astrophysics}

\author{T.C.Weekes}
\address
{ Whipple Observatory, Harvard-Smithsonian CfA,\\ P.O. Box 97,
Amado, AZ 85645-0097, U.S.A.
 }
\maketitle
\begin{abstract}
The development of the atmospheric Cherenkov imaging technique has
led to significant advances in $\gamma$-ray detection sensitivity
in the energy range from 200 GeV to 50 TeV. The Whipple Observatory
10m reflector has detected the first galactic and extragalactic
sources in the Northern Hemisphere; the Crab Nebula has been
established as the standard candle for ground-based $\gamma$-ray
astronomy. The highly variable Active Galactic Nuclei, Markarian
421 and Markarian
501, have proved to be particularly interesting. A new generation
of telescopes with improved sensitivity has the promise of
interesting measurements of fundamental phenomena in physics and
astrophysics. VERITAS (the Very Energetic Radiation Imaging
Telescope Array System) is one such next generation system; it is
an array of seven large atmospheric Cherenkov telescopes planned
for a site in southern Arizona.
\end{abstract}

\section{The Relativistic Universe}

Our universe is dominated by objects emitting radiation via thermal
processes. The blackbody spectrum dominates, be it from the
microwave background, the sun or the
accretion disks around neutron stars. This is the ordinary
universe, in
the sense that anything on an astronomical scale can be considered
ordinary.  It is tempting to think of
the thermal universe as {\it THE UNIVERSE} and certainly it
accounts for much of what we see. However to ignore the largely
unseen, non-thermal, {\it
relativistic}, universe is to miss a major component and one that
is of particular interest to the physicist, particularly the
particle physicist. The relativistic universe is pervasive but
largely unnoticed and involves physical processes that are
difficult to emulate in terrestrial laboratories. 

The most obvious local manifestation of this relativistic universe
is the cosmic radiation, whose origin, 86 years after its
discovery, is still largely a mystery (although it is generally
accepted, {\it but not proven}, that much of it arises in shock
waves
from galactic supernova explosions). The existence of a steady rain
of particles, whose power law spectrum attests to their non-thermal
origin and whose highest energies extend far beyond that achievable
in man-made particle accelerators, attests to the strength and
reach of the forces that power this strange relativistic radiation.
If thermal processes dominate the "ordinary" universe, then truly
relativistic processes illuminate the "extraordinary" universe and
must be studied, not just for their contribution to the universe as
a whole but as the denizens of unique cosmic laboratories where
physics is demonstrated under conditions to which we can only
extrapolate. 

The observation of the extraordinary universe is difficult, not
least because it is masked by the dominant thermal foreground. In
places, we can see it directly such as in the relativistic jets
emerging from AGNs but, even there, we must subtract the foreground
of thermal radiation from the host elliptical galaxy. Polarization
leads us to identify the processes that emit the radio, optical and
X-ray radiation as synchrotron emission from relativistic
particles, probably electrons, but polarization is not unique to
B
synchrotron radiation and the interpretation is not always
unambiguous. The hard, power-law, spectrum of many of the
non-thermal
emission processes immediately suggests the use of the highest
radiation detectors to probe such processes. Hence hard X-ray and
$\gamma$-ray
astronomical techniques must be the observational disciplines of
choice for the exploration of the relativistic universe. Because
the
earth's atmosphere has the equivalent thickness of a meter of lead
for this radiation, its exploitation had to await the development
of space platforms for X-ray and $\gamma$-ray telescopes.

Although the primary purpose of the astronomy of hard photons is
the search for new sources, be they point-like, extended or
diffuse, it opens the door to the investigation of more obscure
phenomenon in high energy astrophysics and even in cosmology and
particle physics. Astronomy at energies up to 10 GeV has made
dramatic progress since the launch of the Compton Gamma Ray
Observatory in 1991 and that work has been summarized
\cite{Gehrels97}.
Beyond 10 GeV it is difficult to efficiently study $\gamma$-rays
from space vehicles, both because of the sparse fluxes which
necessitate large collection areas and the high energies which make
containment a serious problem. The development of techniques
whereby $\gamma$-rays of energy 100 GeV and above can be studied
from the ground, using indirect, but sensitive, techniques is
relatively new and has opened up a new area of high energy photon
astronomy with some exciting possibilities and some preliminary
results. The latter include the detection of TeV
photons from supernova remnants and from the relativistic jets in
AGNs. Such observations seriously constrain the models for such
sources and in many cases lead to the development of a new
paradigm. There remains the possibility that the annihilation lines
from neutralinos might be discovered in the GeV-TeV region, that
the evaporation of primordial black holes might be manifest by the
emission of bursts of TeV photons, that the infrared density of
intergalactic space might be probed by its absorbing effect on TeV
photons from distant sources, and even (in some models) that the
fundamental quantum gravity energy scale might be constrained by
the
observation of short-term TeV flares in extragalactic sources.

\section{Detection Technique}

The techniques of ground-based Very High Energy (VHE) $\gamma$-ray
astronomy are not new but only achieved credibility in the late
eighties with the detection of the Crab Nebula. The most sensitive
technique, the atmospheric Cherenkov imaging technique, is the one
that has been most successful and is now in use at some eight
observatories. Its history and present status has been reviewed
elsewhere \cite{Ong98}. It is an optical "telescope" technique
and thus suffers the usual limitations associated with optical
astronomy: limited duty cycle,
weather dependence, limited field of view. But it also has the
advantage that it is relatively inexpensive because it uses the
same
detector technology (photomultipliers) as optical astronomy, the
same optical reflectors that borrow from solar energy
investigations, and the same pulse processing techniques that are
routinely used in high energy particle physics. In addition the
Cherenkov technique operates in an energy regime where the physics
of particle interactions is relatively well understood and where
there exist advanced Monte
Carlo programs for the simulation of particle cascades.

\begin{figure}
\centerline{\epsfysize 3.5 truein \epsfbox{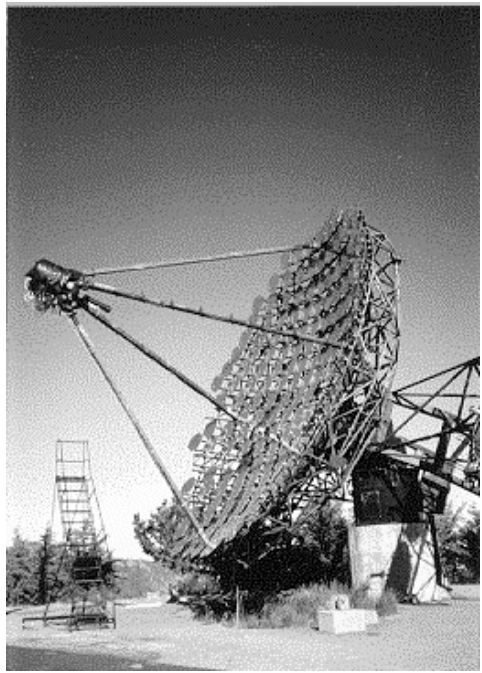}}
\caption{The Whipple Observatory 10m reflector which will be the
prototype for the telescopes in VERITAS.}
\label{10m-fig}
\end{figure}

In recent years, VHE $\gamma$-ray astronomy has been dominated by
two advances in technique: the development of the atmospheric
Cherenkov imaging technique, which led to the efficient rejection
of the hadronic background, and the use of arrays of atmospheric
Cherenkov telescopes to measure the energy spectra of $\gamma$-ray
sources. The former is exemplified by the Whipple Observatory 10m
telescope
(Figure~\ref{10m-fig}) with more modern versions CAT, the French
telescope in the Pyrenees, and the Japanese-Australian CANGAROO
telescope in Woomera, Australia. The most significant examples of
the latter are the five telescope array of imaging telescopes on La
Palma in the Canary Islands which is run by the
Armenian-German-Spanish collaboration, HEGRA, and the four, soon to
be seven, Telescope Array in Utah which is operated by a group of
Japanese institutions. These techniques are relatively mature and
the results from observations with
overlapping telescopes are in good agreement. Vigorous observing
programs are now in progress at all of these facilities; the vital
observing threshold has been achieved whereby both galactic and
extragalactic sources have been reliably detected. Many exciting
results are anticipated as more of the sky is observed with this
generation of telescopes.

\section{Galactic Sources}

It is a measure of the maturity of this new discipline that the
existence and study of galactic sources of TeV radiation is now
considered ordinary and relatively uncontroversial. This is a
dramatic change from only a decade ago when the existence of any
galactic sources at all was hotly contested. These sources were
always variable and difficult to confirm or refute
\cite{Weekes91}; it was not until the observation of steady
sources, in particular, the observation of the Crab Nebula (which
has become the standard candle), that the relative sensitivity of
the different techniques could be assessed and some standards of
credibility set.

The Crab Nebula has been observed by some eight independent groups
and no evidence for variability has been detected. It has been seen
at energies from 200 GeV to more than 50 TeV and accurate energy
spectra have been determined \cite{Hillas98}. Originally predicted
by Gould
\cite{Gould65} as a TeV energy source based on a 
Compton-synchrotron model, the complete $\gamma$-ray spectrum  can
now be fitted by an updated version of the same model
\cite{Hillas98}. The variable parameter in this model is the
magnetic field which is set by the TeV observations at 16$\pm$1 
nanotesla, somewhat smaller than the value estimated from the
equipartition of energy. In practice, recent
optical observations reveal a complex structure at the center of
the nebula (where the TeV photons are believed to originate) and
more sophisticated models are certainly called for.

VHE $\gamma$-rays have also been detected from other galactic
sources. All of these detections are
of sources with negative declinations, best seen in the Southern
 Hemisphere 
where there are fewer VHE observatories and hence the detections
have largely been by one group. The exception is the $\gamma$-ray
pulsar PSR1706-44 which was discovered by the CANGAROO group
\cite{Tanimori95} and confirmed by the Durham group
\cite{Chadwick97}; both of these groups operate from Australia.
The source is detected by EGRET at MeV-GeV energies as 100\%
pulsed. There is no evidence in the TeV signal for pulsations but
there is weak evidence that the pulsar is in a plerion which may be
the source of the TeV $\gamma$-rays. The CANGAROO group also report
the detection of an unpulsed TeV signal from a location close to
the Vela pulsar \cite{Yosh97}; the position coincides with the
birthplace of the pulsar and hence the signal may originate in a
weak plerion left after the ejection of the pulsar. Another
interesting result is the detection of Cen X-3 by the Durham group
\cite{Chadwick98a}.

Perhaps the most surprising (and controversial) result is the
detection of a TeV source that is coincident with one part of the
shell of the supernova remnant, SN1006 \cite{Tanimori98}. 
X-ray observations had shown that there is non-thermal emission
from two parts of the shell
that is consistent with synchrotron emission from
electrons with energy up to 100 TeV; hence the TeV $\gamma$-ray
detection is not a surprise. The TeV emission is consistent with
inverse Compton emission from electrons which have been shock
accelerated in the shell. However it is not clear why it should be
seen from only one region. Because this represents the first direct
detection of SNR shell emission this result, when confirmed, has
great significance. Not only can the magnetic field be estimated
but also the acceleration time; these two parameters are very
important for shock acceleration theory. More sensitive
observations may reveal the detailed energy spectrum, whether or
not the source is extended, and the relative strength of the
TeV emission from each shell. 

\begin{figure} 
\centerline{\epsfysize 2.1 truein \epsfbox{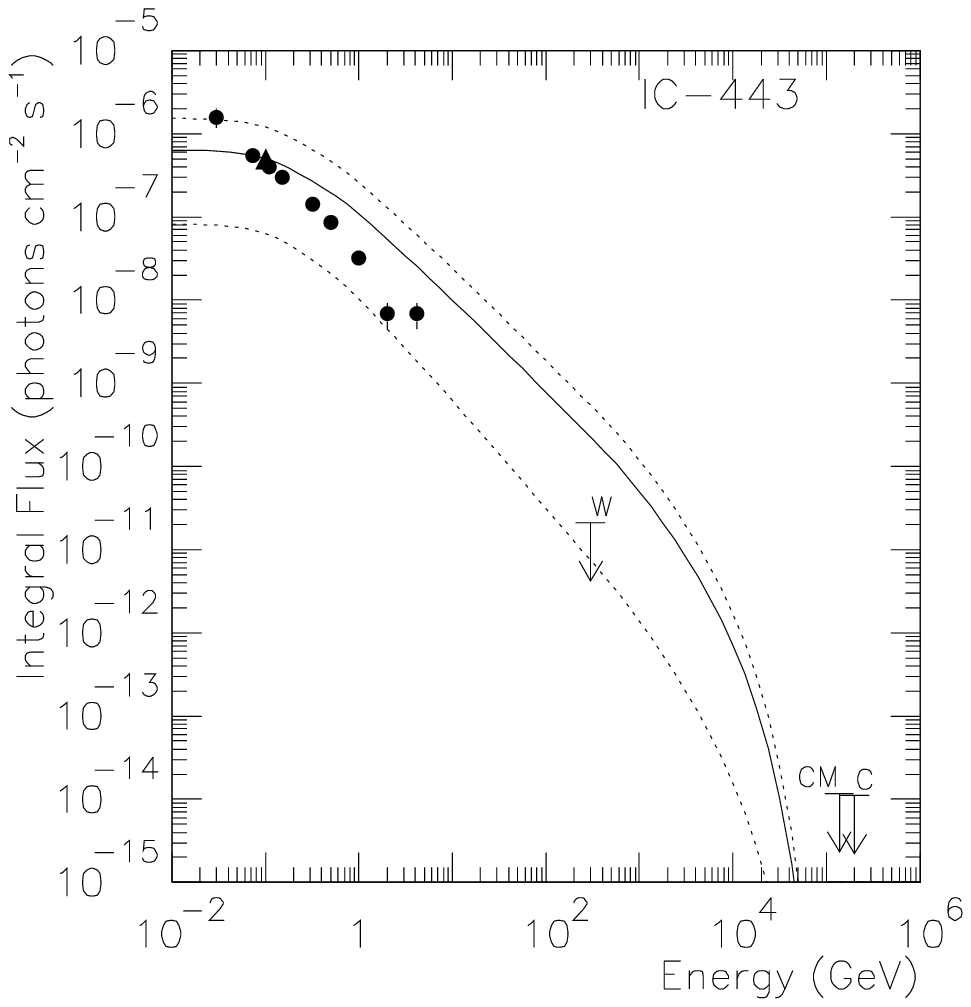}
\epsfysize 2.1 truein \epsfbox{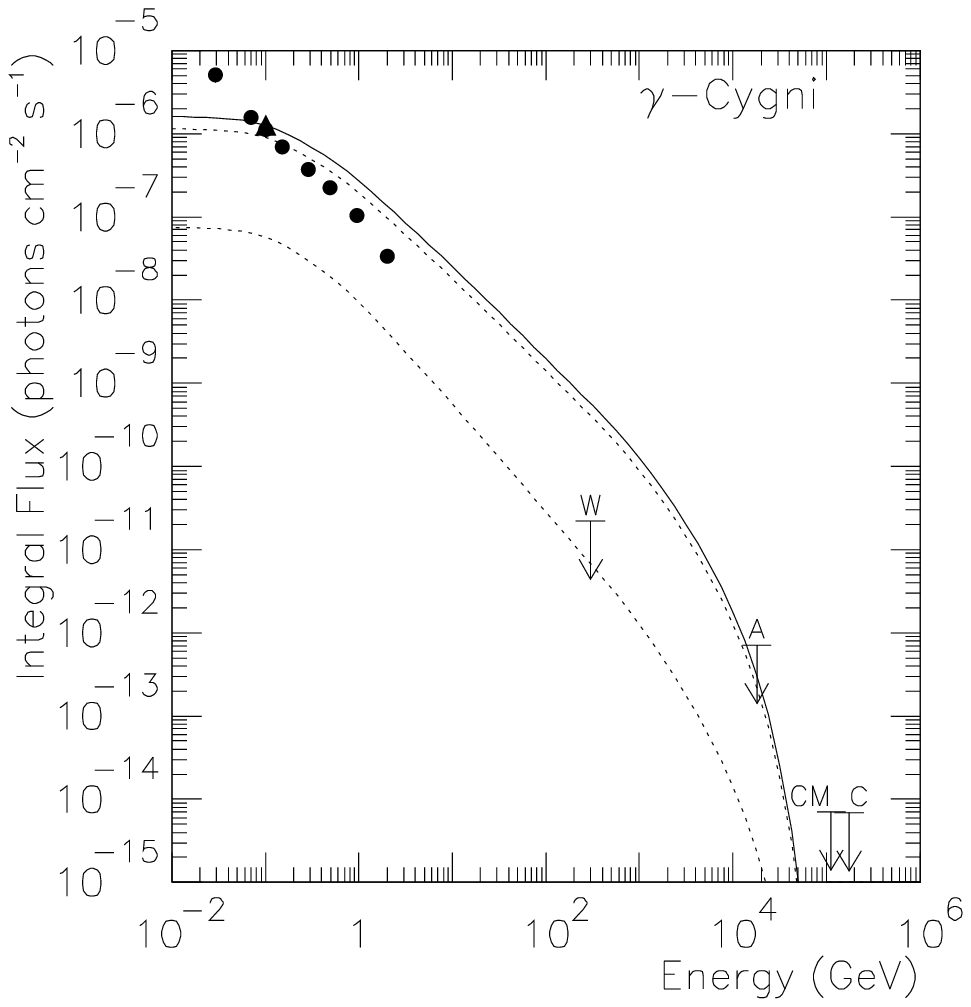}
\epsfysize 2.1 truein \epsfbox{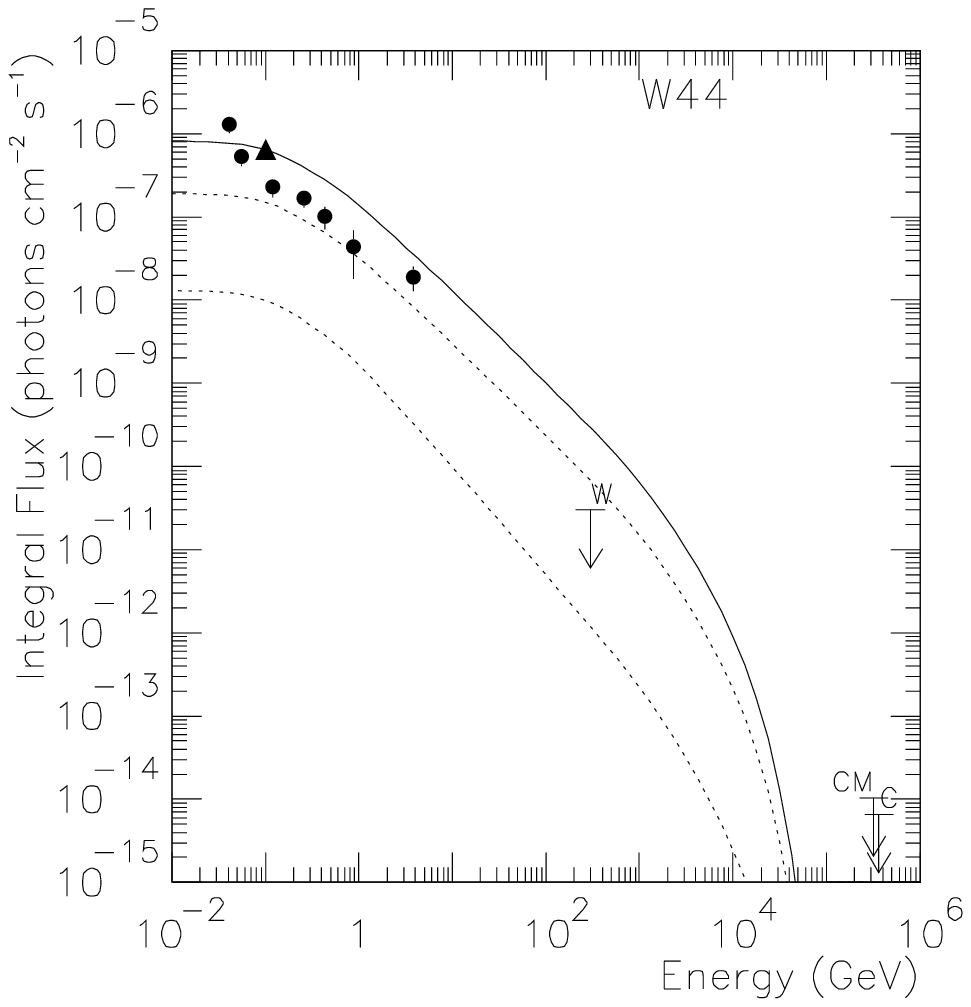}}
\caption{Whipple Observatory flux upper limits (indicated by W) for
shell-type SNRs. The lower energy points are from EGRET. The solid
curves are extrapolations from the EGRET integral fluxes for
$\gamma$-rays arising from $\pi^o$ decay. Upper limits from air
shower arrays are also shown. 
\label{fig3} }
\end{figure}

Ideally, of course, one would like to see direct evidence from VHE
$\gamma$-ray astronomy of emission from hadron collisions in SNR
shells. These SNRs are widely believed to be the source of the
hadronic cosmic rays seen in the solar system (at least up to
proton energies of 100 TeV) which fill the galaxy. However this
canonical model mostly rests on circumstantial evidence and it is
highly desirable to find the smoking gun that would clinch the
issue. Supernovae certainly have sufficient energy and their
occurrence rate is about right; also there is a known mechanism
associated with shock fronts to explain acceleration. Hence when
EGRET detected a small number of $\gamma$-ray sources at GeV
energies which appeared to coincide with known SNRs
\cite{Esposito96}, it was widely believed that the cosmic ray
origin problem had been solved. However Drury et al.
\cite{Drury94} had shown that the $\gamma$-ray spectrum of such
sources should be rather flat power-laws that would extend to TeV
energies. Extensive observations by the Whipple collaboration have
failed to find any evidence for TeV emission \cite{Buckley98}.
The upper limits are shown in Figure~\ref{fig3} along with the
EGRET points. More elaborate models have been constructed
that can be made to fit the observations \cite{Gaisser97}. It is
also possible that the EGRET source/SNR identifications are in
error since the sources are not strong and the galactic 
$\gamma$-ray plane is a confused region at MeV/GeV energies. Either
way, it
would be reassuring for theories of cosmic ray origins to see
definite detections from some shell-type SNRs where the emission is
consistent with $\pi$ production in the shell. The next generation
of VHE detectors should provide these definitive observations.
 
\section{Extragalactic Sources}

\subsection{Relativistic Jets}

\begin{figure}
\begin{minipage}[b]{3.5in}
\centerline{\epsfxsize 3.5 truein \epsfbox{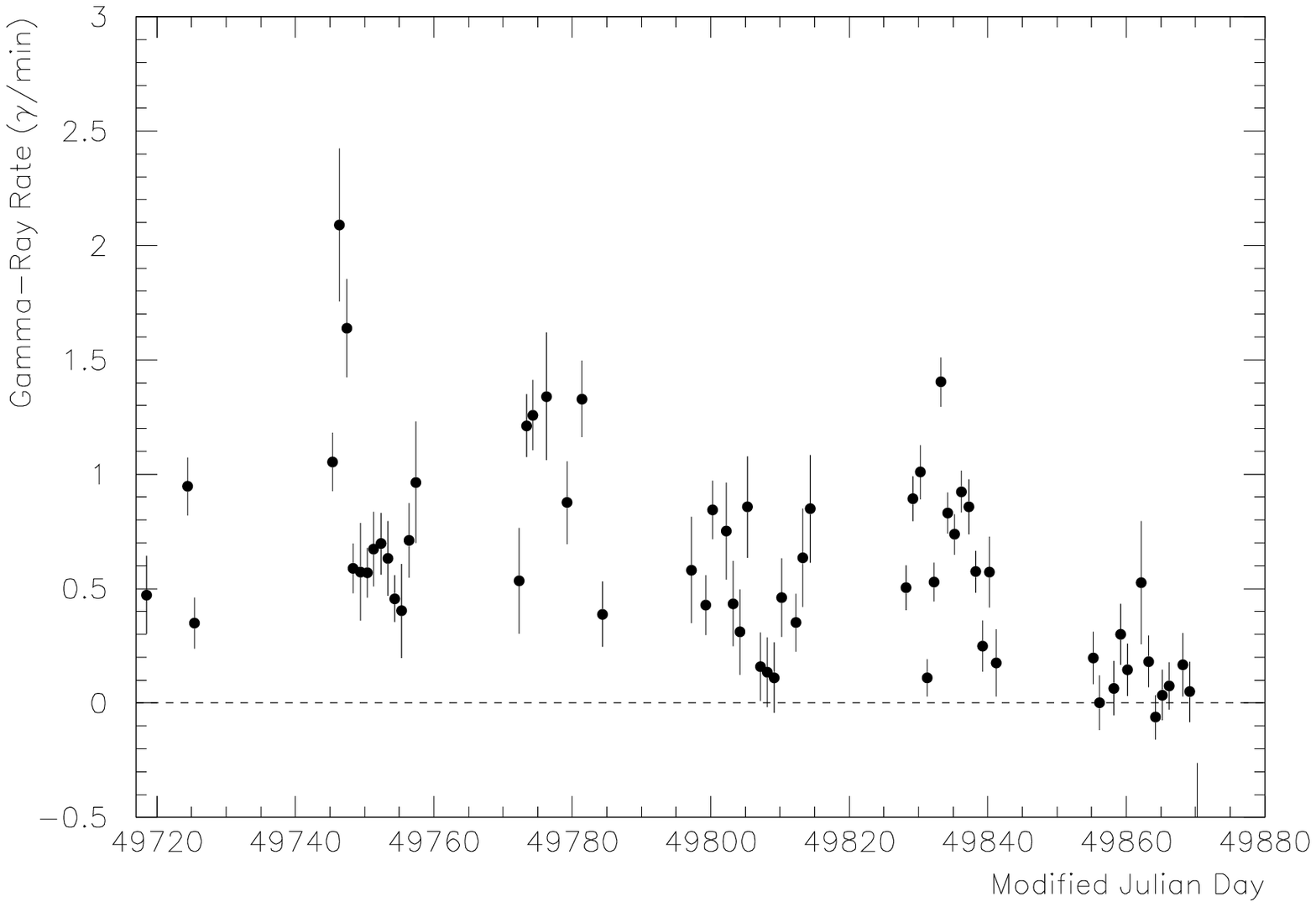}}
\caption{Daily VHE $\gamma$-ray count rates for Mkn 421 during 1995
(from \protect\cite{Buckley96}).}
\label{m4lc}
\end{minipage}
\hfill
\begin{minipage}[b]{3.0in}
\centerline{\epsfxsize 3.0 truein \epsfbox{weekes1406fig4.eps}}
\caption{Average monthly (a) and daily (b) VHE
$\gamma$-ray flux (in units of VHE Crab flux) for Mkn 501 between 
1995 and 1998 (from \protect\cite{Quinn99}).}
\label{m5lc}
\end{minipage}
\end{figure}

One of the most surprising results to come from VHE $\gamma$-ray
astronomy has been the discovery of TeV-emitting blazars. Unlike
the observation of galactic supernovae such as the Crab Nebula,
which
are essentially standard candles, the light-curves of blazars are
highly variable. In Figure~\ref{m4lc} the nightly averages of the
TeV flux from Markarian 421 (Mkn 421) in 1995 are shown as observed
at the Whipple Observatory \cite{Buckley96}. Although AGN
variability was a feature of the AGNs observed by EGRET on the
Compton Gamma Ray Observatory at energies from 30 MeV to 10 GeV,
the weaker signals (because of the finite collection area) do not
allow such detailed monitoring, particularly on short time-scales.

Active galactic nuclei (AGN) are the most energetic on-going
phenomena that we see in extragalactic astronomy. The canonical
model of these objects is that they contain massive black holes
(often at the center of elliptical galaxies) surrounded by
accretion disks and that relativistic jets emerge perpendicular to
the disks; these jets are often the most prominent observational
feature.  Blazars are an important sub-class of AGNs because they
seem to represent those AGNs
which have one of their jets aligned in our direction. Observations
of such objects are therefore unique.

The VHE $\gamma$-ray astronomer is thus in the position of the
particle physicist who is offered the opportunity to observe the
accelerator beam, either head-on or from the side. For the obvious
reason that there is more energy transferred in the forward
direction the particle physicist usually chooses to put his most
important detectors directly in the direction of the beam (or close
to it) and its high energy products. While such observations give
the best insight into the energetic processes in the jet, they do
not give the best pictorial representation. Hence just as it is
difficult to visualize the working of a cannon by looking down its
barrel, it is difficult to get a picture of the jet by looking at
it

 head-on.  Observations at right angles to the jet give us our
best low energy view of the jet phenomenon and indeed provide us
with the spectacular optical pictures of jets from nearby AGNs
(such as M87). 

\subsection{Sources}

Mkn 421 is the closest example of an AGN which is pointing in our
direction. It is a BL Lac object, a sub-class of blazars, so-called
because they resemble the AGN, BL Lacertae which is notorious
because of
the lack of emission lines in its optical spectrum. Because such
objects are difficult, and somewhat uninteresting, for the optical
astronomer they were largely ignored until they were found to be
also strong and variable sources of X-rays and $\gamma$-rays.
Mkn 421 achieved some notoriety largely because it was the first
extragalactic source to be identified as a TeV $\gamma$-ray emitter
\cite{Punch92}. At discovery, its average VHE flux was $\approx$
30\% of the VHE flux from the Crab Nebula. Markarian 501 (Mkn 501),
which is similar to Mkn 421 in many ways, was detected as a VHE
source by the Whipple group in May 1995 \cite{Quinn96}. It was
only 8\% of the level of the Crab Nebula and was near the limit of
detectibility of the technique at that time. The discovery was made
as part of an organized campaign to observe objects that were
similar to Mkn 421 and were at small redshifts. This same campaign
later yielded the detection of the BL Lac object, 1ES 2344+514
\cite{Catanese98a} which is also close (z = 0.044). Recently the
Durham group has announced the detection of the BL Lac object,
PKS2155-304 \cite{Chadwick98b} which is also at a small redshift
(z = 0.116).  
\begin{table}
\caption{Properties of the VHE BL Lac objects}
\label{bllacs-table}
\begin{tabular}{lrccrrr} \hline
  &  & EGRET flux$^a$ & Average flux &  & 
\multicolumn{1}{c}{$\cal{F}_{\rm X}$ $^a$} & 
\multicolumn{1}{c}{$\cal{F}_{\rm R}$ $^a$} \\
\multicolumn{1}{c}{Object} & \multicolumn{1}{c}{z} & (E$>$100 MeV)
& 
(E$>$300 GeV) &  & \multicolumn{1}{c}{(2 keV)} &
\multicolumn{1}{c}{(5 GHz)} \\
  &  & (10$^{-7}$cm$^{-2}$s$^{-1}$) & (10$^{-12}$cm$^{-2}$s$^{-1}$)
& 
\multicolumn{1}{c}{$M_v$ $^a$} & \multicolumn{1}{c}{($\mu$Jy)} & 
\multicolumn{1}{c}{(mJy)} \\ \hline
Mkn 421 & 0.031 & 1.4$\pm$0.2 & 40 & 14.4 & 3.9 & 720 \\
Mkn 501 & 0.034 & 3.2$\pm$1.3  & $\geq$8.1 & 14.4 & 3.7 & 1370 \\
1ES 2344+514 & 0.044 & $<$0.7 & $\leq$8.2 & 15.5 & 1.1 & 220 \\ 
PKS 2155-304 & 0.116 & 3.2$\pm$0.8 & 42 & 13.5 & 5.7 & 310 \\ 
3C 66A & 0.444 & 2.0$\pm$0.3 & 30$^b$ & 15.5 & 0.6 & 806 \\ 
\hline
\multicolumn{7}{p{5.7in}}{$^a$ Radio, optical, and X-ray data from
\cite{Perlman96}.  EGRET data from D.J. Thompson (priv. comm.), 
\cite{Mukherjee97}, and \cite{Kataoka98}.} \\
\multicolumn{7}{l}{$^b$ 1 TeV flux value.}
\end{tabular}
\end{table}
A more controversial, but potentially more
important
detection, is that of 3C 66A reported by the Crimean group
\cite{Neshpor98}. These sources are summarized in
Table~\ref{bllacs-table}. 
Whereas the first two sources have been
seen by a number of groups, the last three are reported by only one
group and require confirmation.

\subsection{Variability}

Perhaps the most exciting aspect of these detections is the
observation of variability on time-scales from minutes to hours.
The very large collection areas ($> 10,000 m^{2}$) associated with
atmospheric Cherenkov Telescopes is ideally suited for the
investigation of short term
variability. The VHE emission from the two best observed sources,
Mkn 421
and Mkn 501 (Figure~\ref{m5lc}), varies by a factor of a hundred.
Although
many hundreds of hours have now been devoted to their study, the
variations are so complex that it is still difficult to
characterize their emissions. It has been suggested
\cite{Buckley96} that for Mkn 421
the emission is consistent with a series of short flares above a
baseline that falls below the threshold of the Whipple telescope
(Figure~\ref{m4lc}); the average flare duration is one day or
shorter.

\begin{figure}
\centerline{\epsfysize 2.5 truein \epsfbox{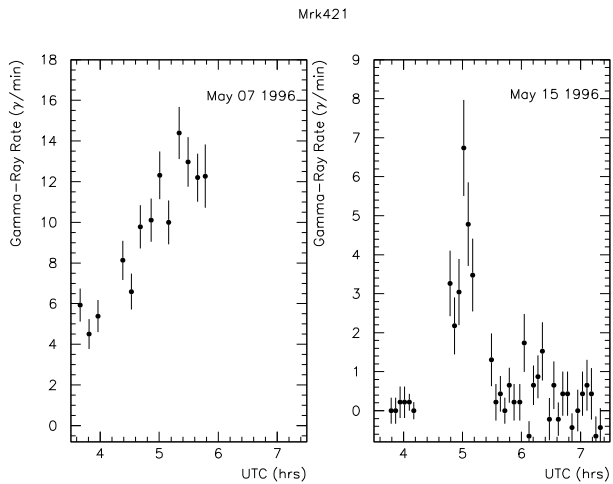}}
\caption{Mkn 421 flares of 1996 May 7 (left) and
May 15 (right) (adapted from \protect\cite{Gaidos96}).}

\label{bigones}
\end{figure}

The most important observations of Mkn 421 were in May, 1996 when
it was found to be unusually active. On May 7, a flare was observed
with the largest flux ever recorded from a VHE source. The
observations began when the flux was already several times that of
the Crab Nebula and it continued to rise over the next two hours
before levelling off (Fig.~\ref{bigones}). Observations were
terminated as the moon rose but the following night it was observed
at its quiescent level. One week later (May 15) a smaller, but
shorter, flare was detected; in this case the complete flare was
observed and the doubling time in the rise and fall was $\approx$
15 minutes. This is the shortest time variation seen in any
extragalactic $\gamma$-ray source at energies $>$ 10 MeV (apart
from in a $\gamma$-ray burst).

Mkn 501 is also variable, but as at other
wavelengths, the characteristic time seems longer. Its baseline
emission has varied by a factor of 15 over four years
\cite{Quinn99} (Figure~\ref{m5lc}). Hour-scale variability has also
been detected but its most important time variation characteristic
appears to be the slow variations seen over the five months in
1997.

\subsection{Spectrum}

The atmospheric Cherenkov signal is essentially calorimetric and
hence it should be possible to derive the $\gamma$-ray energy
spectrum from the observed light pulse spectrum. In practice it is
more difficult because, unless an array of detectors is used, the
distance to the shower core (impact parameter) is unknown. Although
the extraction of a spectrum from even a steady and relatively
steady source as the Crab Nebula required considerable effort and
the development of new techniques, it was relatively easy to
measure
the spectra of Mkn 421 and Mkn 501 in their high state because the
signal was so strong. The general features of the spectra derived
from the Whipple observations are in agreement with those derived
at the HEGRA telescopes \cite{Lorenz98}.

\begin{figure}
\centerline{\epsfysize 2.5 truein \epsfbox{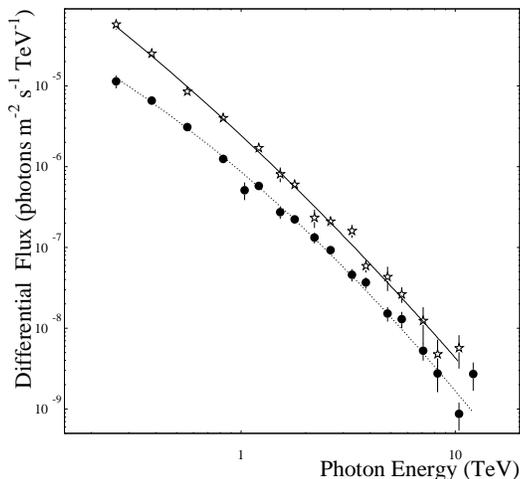}}
\caption{VHE spectra of Mkn 421 (filled circles) and
Mkn 501 (open stars) as measured with
the
Whipple Observatory telescope \protect\cite{Krennrich98}. 
\label{tevspecs}
}
\end{figure}

The May 7, 1996 flare of Mkn 421 provided an excellent data base
for the extraction of a spectrum; the data can be fit by a simple
power-law $(dN/dE \propto E^{-2.6})$. There is no evidence of a
cutoff up to energies of 5 TeV \cite{Zweerink97}
(Figure~\ref{tevspecs}). Because of the possibility of a high
energy
cutoff due to intergalactic absorption there is considerable
interest in the highest energy end of the spectrum. Large zenith
angle observations at Whipple \cite{Krennrich97} and observation
by HEGRA \cite{Lorenz98} confirm the absence of a cutoff out to
10 TeV.

The generally high state of Mkn 501 throughout 1997 give data from
the Whipple telescope that can be best fit by a curved spectrum of
the form: $dN/dE$ and $E^{-2.20-0.45log_{10}E}$ \cite{Samuelson98}
(Figure~\ref{tevspecs}). Here the spectrum extends to at least 10
TeV. The curvature in the spectrum could be caused by the intrinsic
emission mechanism or by absorption in the source. Since Mkn 421
and Mkn 501 are virtually at the same redshift it is unlikely that
it could be due to intergalactic absorption since Mkn 421 does not
show any curvature \cite{Krennrich98}.

\subsection{Multiwavelength Observations}

The astrophysics of the $\gamma$-ray emission from the jets of AGNs
are best explored using multiwavelength observations. These are
difficult to organize and execute because of the different
observing constraints on radio, optical, X-ray,

space-based
$\gamma$-ray
and ground-based $\gamma$-ray observatories. Of necessity
observations are often incomplete and, when complete coverage is
arranged, the source does not always cooperate by behaving in an
interesting way! 

\begin{figure}
\centerline{\epsfysize 3.5 truein \epsfbox{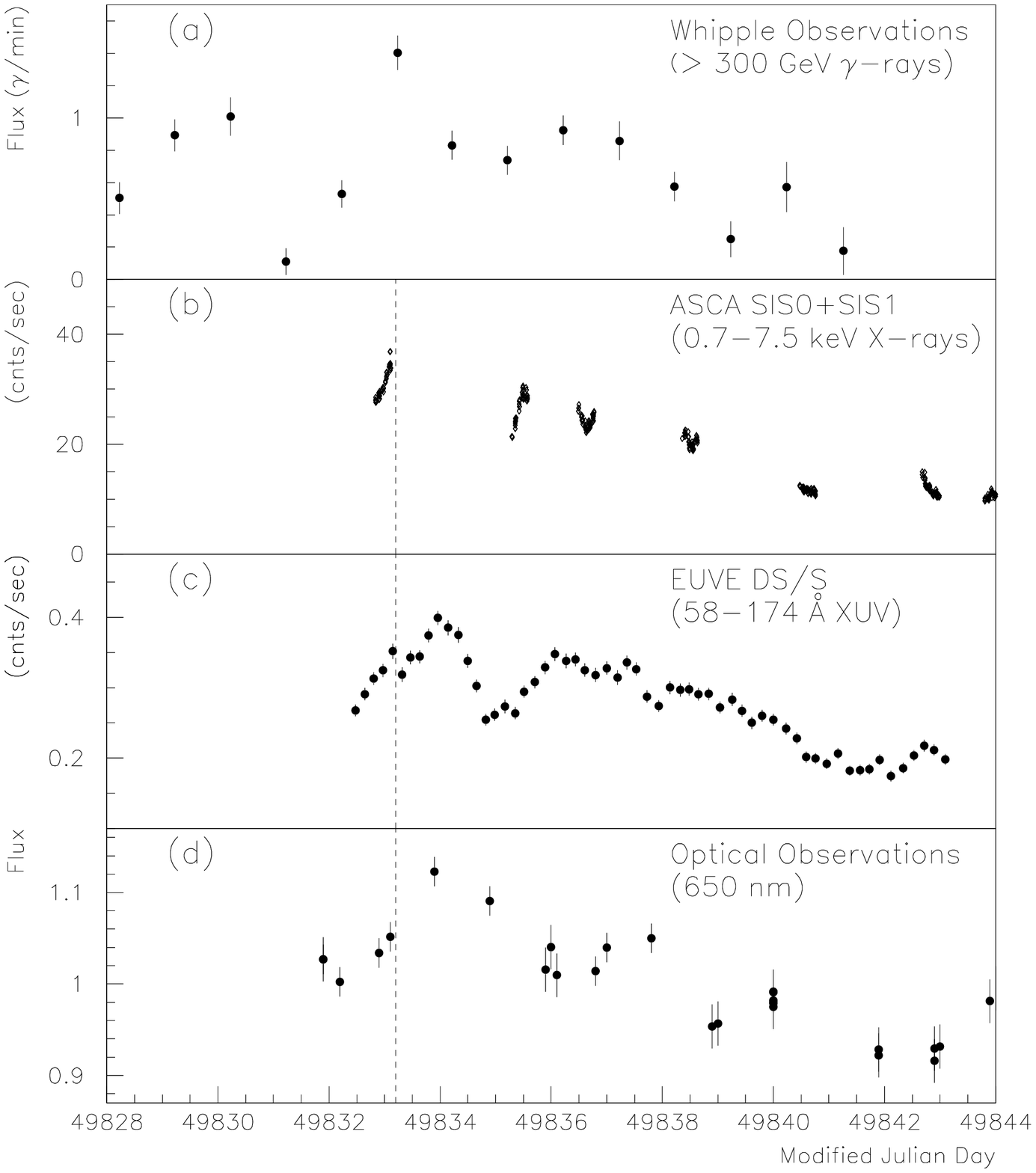}
 \epsfysize 3.5 truein \epsfbox{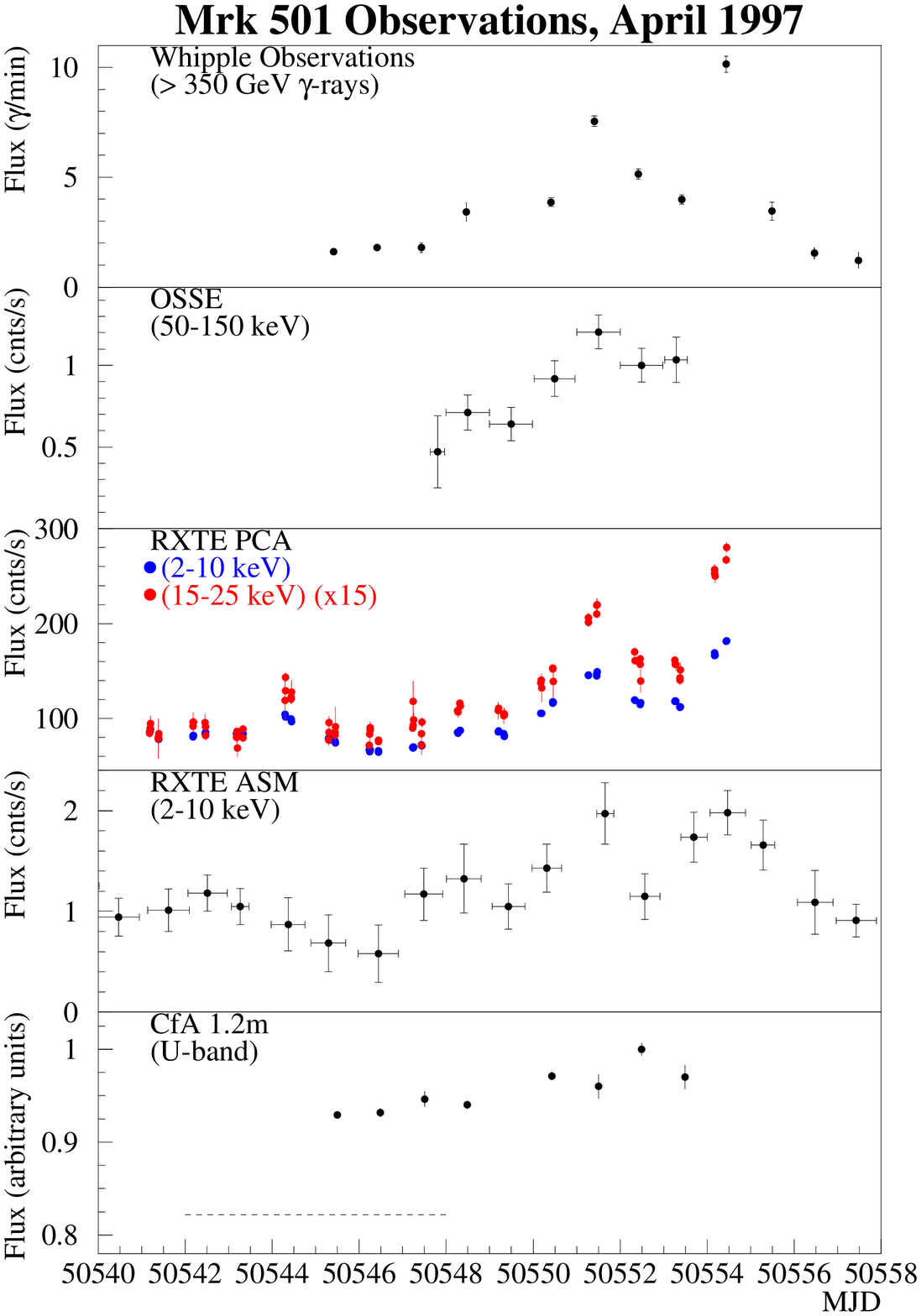}}
\caption{{\it Left}: Multi-wavelength observations of Mrk 421 (from
\protect\cite{Buckley96}): (a) VHE $\gamma$-ray, (b) X-ray, (c)
extreme UV, and (d) optical lightcurves taken during the period
1995 April-May (April 26 corresponds to MJD 49833).  {\it Right:}
Multi-wavelength observations of Mkn 501 (adapted from
\protect\cite{Catanese97}): (a) $\gamma$-ray, (b) hard X-ray, (c) soft
X-ray, (d) U-band optical taken during the period 1997 April 2-20
(April 2 corresponds to MJD 50540).  The dashed line in (d)
indicates the optical flux in 1997 March.}
\label{multilc}

\end{figure}

The first multiwavelength campaign on Mkn 421 coincided with a TeV
flare on May 14-15, 1994 and showed some evidence for correlation
with the X-ray band; however no enhanced
activity was seen in EGRET \cite{Macomb95}. A year later, in a
longer campaign, there was again correlation between the TeV flare
and the soft X-ray and UV data but with an apparent time lag of the
latter by one day \cite{Buckley96} (Figure~\ref{multilc}). The
variability amplitude is comparable in the X-ray and TeV emission
($\approx$ 400\%) but is smaller in the EUV ($\approx$200\%) and
optical ($\approx$20\%) bands. In April, 1998 there was again a
correlation seen between an X-ray flare observed by SAX and
Whipple; in this case the TeV flare was much shorter (a few hours)
compared to the X-ray (a day) \cite{Maraschi98}.

The first multiwavelength campaign on Mkn501 was undertaken when
the TeV signal was seen to be at a high level. The surprising
result was that the source was detected by the OSSE experiment on
CGRO in the 50-150 kev band (Figure~\ref{multilc}). This was
the highest flux ever recorded by OSSE from any blazar (it has not
detected Mkn 421) but the amplitude of the X-ray variations
($\approx$200\%) was less than those of the TeV $\gamma$-rays
($\approx$400\%) \cite{Catanese97}.

\subsection{Multiwavelength Power Spectra}

Because of the strong variability in the TeV blazars it is
difficult to represent their multiwavelength spectra. In
Figure~\ref{multiwave-fig} we show the fluxes plotted as power
($\nu \times F_{\nu}$) from Mkn 421 and Mkn 501 during flaring as
well as
the average fluxes. Both sources display the two peak distribution
characteristic of Compton-synchrotron models, e.g., the Crab
Nebula.
Whereas the synchrotron peak in Mkn 421 occurs near 1 keV, that of
Mkn 501 occurs beyond 100 keV which is the highest seen from any
AGN. In 1998 the synchrotron spectrum peak in Mkn 501 shifted back
to 5 keV and the TeV flux fell below the X-ray flux. 

\subsection{Implications}

The sample of VHE emitting AGNs is still very small but it is
possible to draw some conclusions from their properties (summarized
in Table~\ref{bllacs-table}).

\begin{itemize}

\item The first three objects, all detected by the Whipple group,
are the three closest BL Lacs in the northern sky. Some 20 other BL
Lacs  have been observed with z $<$ 0.10 without detectable
emission. This could be fortuitous,  because they are standard
candles and these are closest (but the distance differences are
small), or because they suffer the least absorption (but there is
no cutoff apparent in their spectra).

\item All of the objects are BL Lacs; because such objects do not
show emission lines and therefore
probably do not have strong optical/infrared absorption close to
the source, it is suggested that BL Lacs are preferentially VHE
emitters.
\begin{figure}
\centerline{\epsfysize 2.7 truein \epsfbox{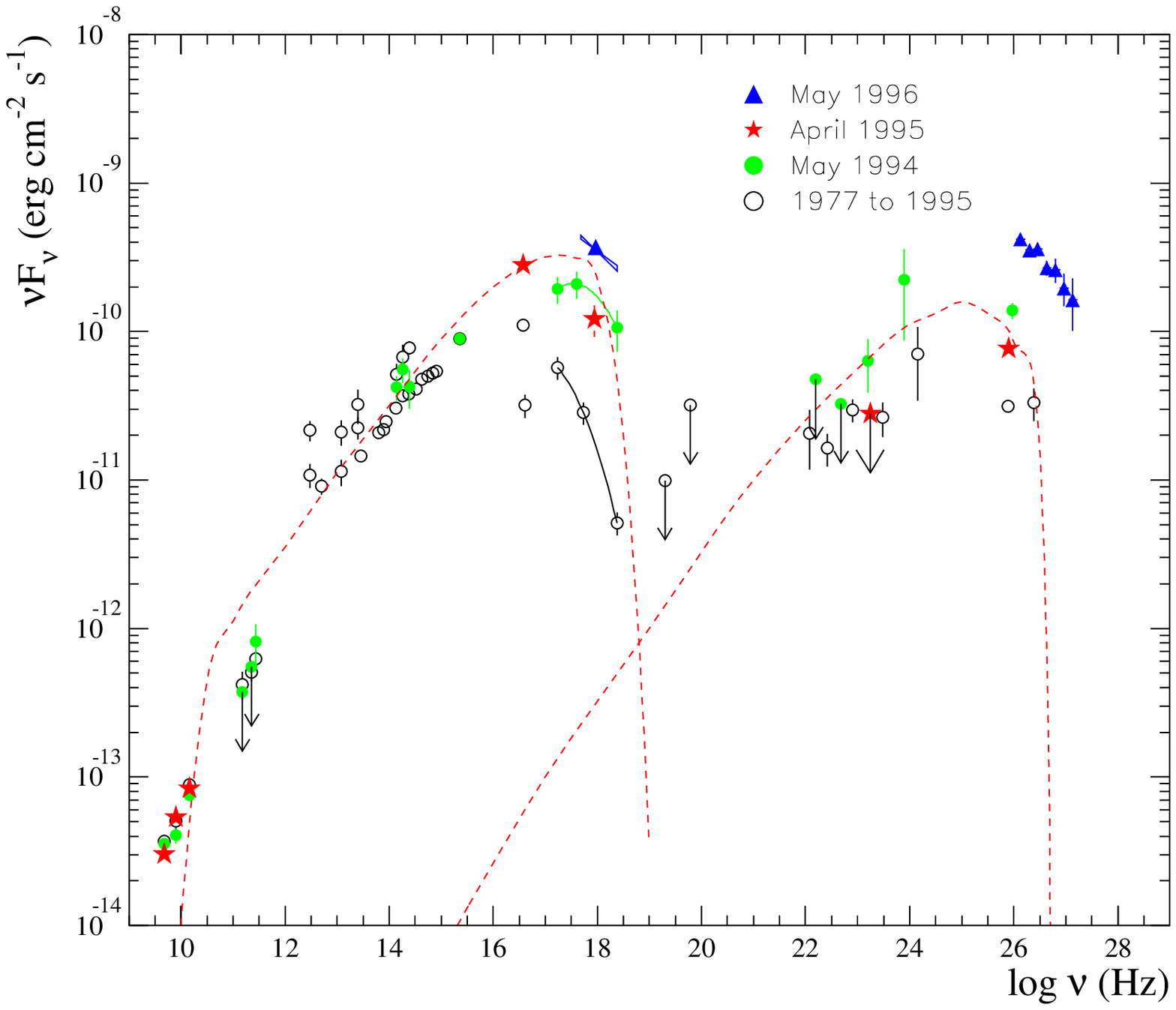}
\epsfysize 2.7 truein \epsfbox{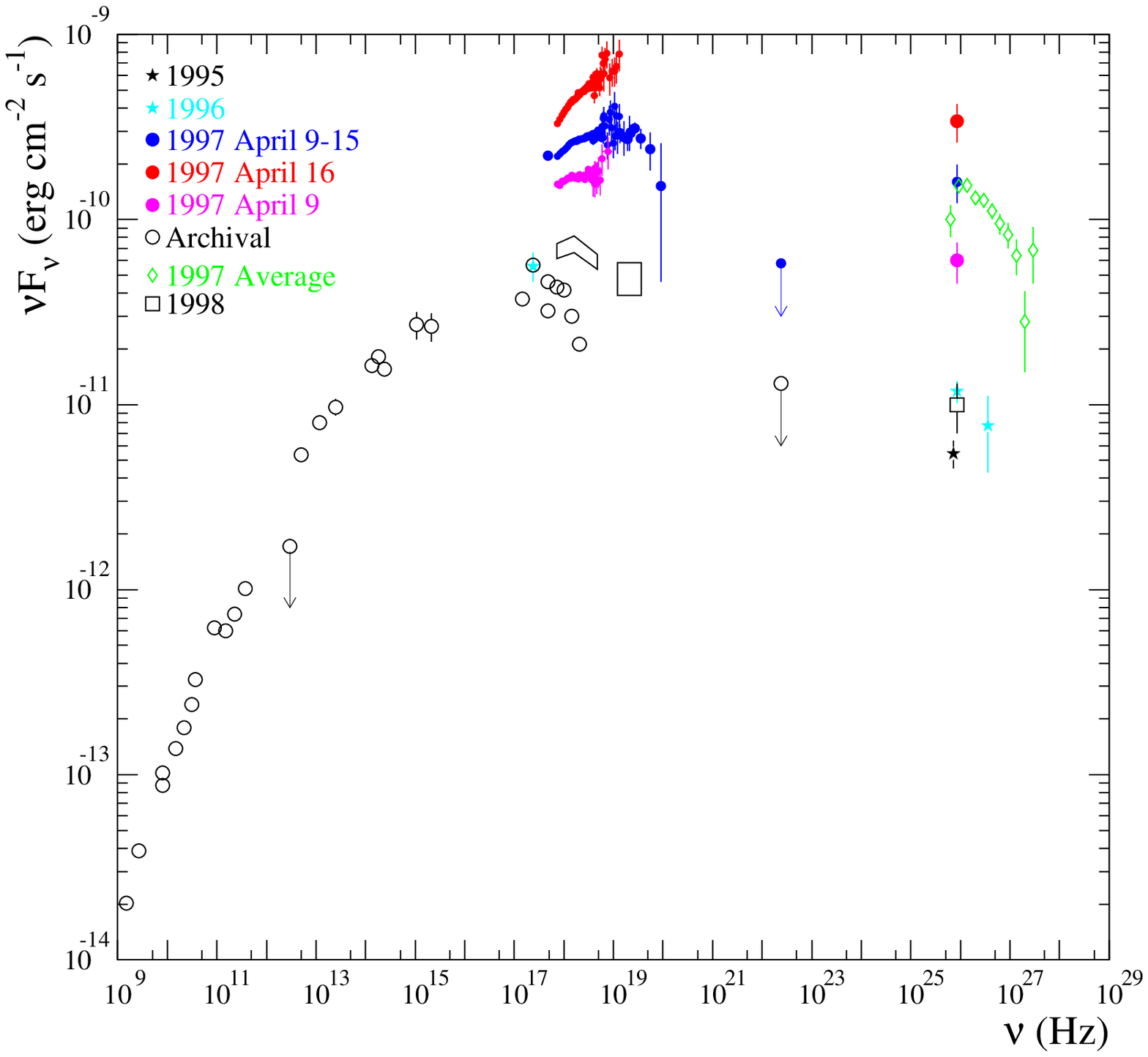}}
\caption{{\it Left:} The multi-wavelength power spectrum of Mkn 421
(adapted from \protect\cite{Buckley96}).  The dashed line shows an
SSC
model fit to the data.  {\it Right:} The multi-wavelength power
spectrum of Mkn 501 (adapted from \protect\cite{Catanese97}).
\label{multiwave-fig}
}
\end{figure}

\item Four of the five sources are classified as XBLs which
indicates that they are strong in the X-ray region and that the
synchrotron spectrum most likely peaks in that range (and that the
Compton spectrum peaks in the VHE $\gamma$-ray range). The fifth,
3C 66A, is an RBL, like many of the blazars detected by EGRET; it
is believed that these blazars have synchrotron spectra that peak
at lower energies and Compton spectra that peak in the HE 
$\gamma$-ray region.

\item Only three (Mkn 421, PKS 2155-304 and 3C 66A) are listed in
the Third EGRET Catalog; there is a weak detection reported by
EGRET for Mkn 501.

\item If 3C 66A is confirmed (and to a lesser extent PKS 2155-305),
then the intergalactic absorption is significantly less than had
been suggested from galactic evolution models. 

\item There is evidence for variability in all of the sources. The
rapid variability seen in Mkn 421 indicates that the emitting
region is very small which might suggest it is close to the black
hole. In that case the local absorption must be very low (low
photon
densities). It seems more likely that the region is well outside
the dense core.

\end{itemize}

There are three basic classes of model considered to explain the
high energy properties of BL Lac jets: Synchrotron Self Compton
(SSC), Synchrotron External Compton (SEC) and Proton Cascade (PC)
Models. In the first two the progenitor particles are electrons, in
the third they are protons.  VHE $\gamma$-ray observations have
constrained the types of models that are likely to produce the
$\gamma$-ray emission but still do not allow any of them to be
eliminated.  For instance, the correlation of the X-ray and the VHE
flares is consistent with the first two models where the same
population of electrons radiate the X-rays and $\gamma$-rays. 
There is little evidence for the IR component in BL Lac objects
which would be necessary in the SEC models as the targets for
Compton-scattering, so this particular type of model may not be
likely
for these objects.  The PC models which produce the $\gamma$-ray
emission through $e^+e^-$ cascades also have great difficulty
explaining the rapid cooling observed in the TeV emission from 
Mkn~421. Also the high densities of unbeamed photons near the
nucleus, such as the accretion disk or the broad line region, are
required to initiate the
cascades and these cause high pair opacities to TeV $\gamma$-rays
\cite{Coppi93}.  

Significant information comes from the multiwavelength campaigns
(although thus far these have been confined to Mkn 421 and Mkn
501). Simultaneous measurements constrain the magnetic field
strength ($B$) and Doppler factor ($\delta$) of the jet when the
electron cooling is assumed to be via synchrotron losses.  The
correlation between the VHE $\gamma$-rays and optical/UV photons
observed in 1995 from Mkn 421 indicates both sets of photons are
produced in the same region of the jet; $\delta \gsim 5$ is
required for the VHE photons to escape significant pair-production
losses \cite{Buckley96}.  If the VHE $\gamma$-rays are produced
in the synchrotron-self-Compton process, $\delta = 15 - 40$ and $B
= 0.03 - 0.9$G for Mrk 421 \cite{Catanese98b}, \cite{Tavecchio98}
and $\delta < 15$ and $B = 0.08 - 0.2$G for Mkn 501
\cite{Samuelson98}, \cite{Tavecchio98}. On the other hand by
assuming protons produce the $\gamma$-rays in
Mkn 421, Mannheim \cite{Mannheim93} derives $\delta = 16$ and $B
= 90$G.  The Mkn 421 values of $\delta$ and $B$ are extreme for
blazars, but they are still within allowable ranges and are
consistent with the extreme variability of Mkn 421.

\section{Intergalactic Absorption}

Thus far it has not been possible to make a direct measurement of
the infrared background radiation at wavelengths more than 3.5
microns and less than 140 microns. This is unfortunate since the
background potentially
contains valuable information for cosmology, galaxy formation and
particle
physics. The problem for direct measurement is the presence of
foreground local and galactic sources. However the infrared
background can make its presence felt by the absorption it produces
on the spectra of VHE $\gamma$-ray sources when they are at great
distances. The absorption is via the $\gamma \gamma \rightarrow
e^+e^-$
process, the physics of which is well understood. The maximum
absorption occurs when the product of the energy of the two photons
($\gamma$-ray and infrared) is approximately equal to the product
of the rest masses of the electron-pair. Hence a 1 TeV $\gamma$-ray
is most heavily absorbed by 0.1eV (1.2 micron) infrared photon in
head-on collisions. 

The importance of this effect for VHE and UHE $\gamma$-ray
astronomy was first pointed out by Nikishov \cite{Nikishov62};
its potential for making an indirect measurement of the infrared
background was pointed out by Gould and Schreder \cite{Gould67}
and, more
recently, in the aftermath of the EGRET detections of AGNs, by
Stecker and de Jager \cite{Stecker93}. At the redshift of the AGNs
detected at VHE
energies to date (0.03 to 0.5) if the infrared density has the
value assumed in some models \cite{Stecker93}, the effect is
appreciable and should be apparent in carefully measured energy
spectra in the range 1 to 50 TeV. 

\begin{figure}[t!]
\centerline{\epsfysize 2.5 truein \epsfbox{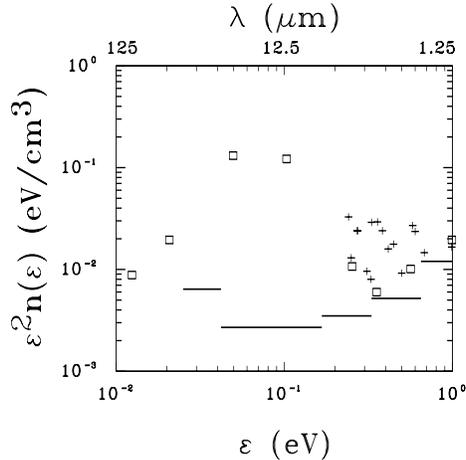}}
\caption{Upper limits to the background IR density.  The solid
lines show the limits derived assuming the VHE spectra of Mkn 421
and Mkn
501 extend to 10 TeV and the dashed lines show the limits if the
spectra extend only to 6 TeV.  Other symbols represent upper limits
from direct measurements (squares are from COBE/DIRBE
measurements).
Figure from (\protect\cite{Biller98a}).
\label{main-irlim-fig}
}
\end{figure}

Ideally for such a measurement the intrinsic emission spectrum of
the
$\gamma$-rays from the distant source should be known. In practice
this is not the case although thus far all the AGNs detected in the
GeV-TeV range appear to have very smooth power-law spectra. Biller
et al. \cite{Biller98a} have made a conservative derivation of
upper limits on the infrared spectrum based on the measured
$\gamma$-ray spectrum from 0.5 to 10 TeV from Mkn 421 and  Mkn 501
by the Whipple and HEGRA groups. These upper
limits apply to
infrared energies from 0.025 to 1.0 eV; they are the best upper
limits over this range. At some wavelengths, these limits are as
much as an
order of magnitude below the upper limits set by the DIRBE/COBE
satellite (see Figure~\ref{main-irlim-fig}).

The infrared densities are calculated such that they do not cause
the shape of the observed VHE spectrum to deviate from the bounds
set from the VHE measurements. This approach has the effect of
anchoring the lower energy TeV data to the appropriate infrared
upper limits and then extending these bounds so that they are
consistent with those based on the shape and extent of the AGN
spectra at the higher energies. Thus the maximum energy density in
each interval of infrared energy is determined; these limits are
plotted in Figure~\ref{main-irlim-fig} where a 
maximum energy of 10 TeV is considered; also shown are the
upper limits from other methods.

These upper limits do not conflict with the predictions of the
infrared background based on detailed models of galactic evolution
\cite{Primack98}. They do however allow some more cosmological
possibilities to be eliminated. In particular in one scenario,
density fluctuations in the early universe (z $\approx$ 1000) could
have produced very massive objects which would collapse to black
holes at later times and could explain the dark matter. However
although undetectable now, they would have produced an amount of
infrared radiation that would have exceeded the above limits
\cite{Biller98a}. These limits also place some constraints on
radiative neutrino decay.

\section{Gamma Ray Bursts}

The contribution of TeV observations to the physics of $\gamma$-ray
bursts is
at once the most speculative and most important (potentially) of
all the scientific topics considered here. As yet, there is no
positive detection of TeV photons during or immediately after a
classical $\gamma$-ray burst (GRB) (although there is one
tantalizing but unconfirmed observation \cite{burst98}). However
since there is no turn over seen in the spectra of GRBs detected by
EGRET at energies $>$ 30 GeV, there is the potential for
interesting observations at VHE energies. The observed EGRET
spectra are power laws with differential indices 1.95${\pm}$0.25
\cite{Dingus98}. The sensitivity of current ACTs is such that
sources with spectral indices $\approx$ 2 would be easily
detectable even for fluences as low as $5 \times 10 ^{-8}$
ergs/cm$^{2}$ \cite{Connaughton96}. Although only four of the
very bright BATSE bursts were seen by EGRET, these were the
brightest to occur within the field of view of EGRET and there is
nothing to suggest that all bursts might not have GeV-TeV
components. In fact, EGRET was not a very sensitive detector for
GRBs both because of its limited collection area and its deadtime.
There are now several models that suggest that TeV emission may be
a strong feature of GRBs \cite{Totani98}, \cite{Dermer98}.

There are however several negative factors concerning the possible
detection of GRBs by ACTs. The narrow field of view combined with
the low
duty-cycle (clear, dark nights) lessens the chance of the
serendipitous detection of the TeV component of a GRB. If the GRBs
are truly cosmological (as they appear to be), then intergalactic
absorption by pair production on infra-red photons must come into
play at some point, steepening the apparent spectra. However the
next generation of ACTs will have reduced energy thresholds, better
flux sensitivities and rapid slew capabilities; these features
combined with the more accurate source locations anticipated with
the launch of HETE-2 may provide TeV detections at the rate of a
few per year. In addition, the water Cherenkov detector, MILAGRO,
will have all sky coverage (although reduced sensitivity below 1
TeV) and will have guaranteed coverage of some bursts detected by
satellites.

A feature of the EGRET GRB observations was that there was evidence
for delayed emission (up to 1.5 hours) from the burst site
\cite{hurley94}. This may indicate a different component at these
energies. Some models \cite{Totani98} predict that this delayed
emission could persist for days and could hence be easily observed
with narrow field of view instruments.

The detection of a TeV $\gamma$-ray component in a GRB would be a
serious parameter for the emission models, in particular the
Lorentz
bulk motion in the source would be constrained. It would also be an
independent distance indicator since the source would have to show
absorption if the redshift was $>$ 0.1.

\section{Neutralinos}

The best candidate for the cold dark matter component is the
neutralino, the lightest supersymmetric particle. These particles
annihilate with the emission of a single $\gamma$-ray line whose
energy is equal to that of the neutralino mass; however other
annihilation modes
are possible and there may be a $\gamma$-ray continuum.
 There are limits on the possible
masses from cosmology and from accelerator experiments but the
range from 30 GeV to 3 TeV is allowed. The upper part of this range
would be accessible for study by ground-based $\gamma$-ray
telescopes.
The neutralinos, if they exist, would be expected to cluster to the
center of galaxies and might be detectable by their $\gamma$-ray
emission, either as a line or a continuum. Detailed numerical
simulations indicate that there may be a strong density enhancement
towards the centers of galaxies such as our own. Hence the Galactic
Center is a prime candidate for observations. This hypothesis is
given some credence by the detection of a somewhat extended
$\gamma$-ray source at the Galactic Center \cite{mayer98} at
energies above 300 MeV.

In a recent paper, Bergstrom, Ullio and Buckley
\cite{Bergstrom98} have estimated the flux from the annihilation
radiation of neutrinos in the Galactic Center using the most recent
models of the galactic mass distribution. The predicted line has a
relative
width of 10$^{-3}$. Neither space nor ground-based detectors have
energy resolution of this quality (even in the next generation of
detectors) but the intensity of the line is such that it might be
detectable even with relatively crude energy resolution.

\section{Quantum Gravity}

Some quantum gravity models predict the refractive index of light
in vacuum to be dependent on the energy of the photon. This effect,
originating from the polarization of space-time, causes an energy-
dependance to the velocity of light.
Effectively, the quantum fluctuations are on distance scales near
the
Planck length, ($L_P \simeq 10^{-33}$cm), (corresponding to 
time-scales of 1/$E_P$, the Planck mass ($\simeq
10^{19}$GeV)). Different models of quantum gravity give widely
B
differing predictions for the amount of time dispersion. In one
model the first order time dispersion is given by:
\begin{equation}
\Delta t \simeq \xi \frac{E}{E_{QG}} \frac{L}{c}
\label{main-qg-eq}
\end{equation}
where $\Delta t$ is the time delay relative to propagation at the
velocity of light, $c$,  $\xi$ is a model-dependent factor of order
1, $E$ is the energy of the observed photons, $E_{QG}$ is the
quantum energy scale, and $L$ is the distance from the source. In
most models $E_{QG} \approx E_{P}$ but, in recent work in the
context of string theory, it can be as low as 10$^{16}$GeV
\cite{Witten96}. 

Recently it has been suggested that astrophysical observations of
transient high energy emission from distant sources might be used
to measure (or limit) the quantum gravity energy scale. 
Amelino-Camelia et al. \cite{Amelino98} suggested that BATSE
observations
of GRBs would provide a powerful method of probing this fundamental
constant if variations on time-scales of milliseconds could
be measured in the MeV signal in a GRB which was measured to be at
a cosmological distance. Such time-scales and distances have been
measured in GRBs but so far not in the same GRB. The absence of
time dispersion in flares of TeV $\gamma$-rays from AGNs at known
distances provides an even more sensitive measure. Biller et al.
\cite{Biller98b} have used the sub-structure observed in the 15
minute flare in Mkn 421 observed by the Whipple group on April 15,
1996 \cite{Gaidos96} to derive a lower limit on $E_{QG}$.

On a time-scale of 280 seconds there is weak (2$\sigma$) evidence
for correlated variability in two energy ranges: 300 GeV to 2 TeV
and $>$ 2 TeV. For a Hubble Constant of 85 km/s/Mpc, the distance
$L$ is $1.1 \times10^{16}$ light-seconds. This gives a lower limit
to $E_{QG}$ of $>4 \times 10^{16}$GeV assuming $\xi$ is $\approx$
1. This is the most convincing lower limit on $E_{QG}$ to date.

Because VHE $\gamma$-ray astronomy is still in its infancy and the
exposure time on AGNs still limited, it is likely that much more
sensitive measurements will lead to better limits on $E_{QG}$ as a
new generation of detectors comes on-line and permits the detection
of shorter time-variations and/or more distant sources. 

\section{Future Prospects}

It is clear that to fully exploit the potential of 
ground-based $\gamma$-ray astronomy the detection techniques
must be improved. This will happen by extending the energy coverage
of the technique and by increasing its flux sensitivity. Ideally
one would like to do both but in practice there must be trade-offs.
Reduced energy threshold can be achieved by the use of larger but
cruder mirrors and this approach is currently being exploited using
existing arrays of solar heliostats (STACEE and CELESTE). A 
German-Spanish project (MAGIC) to build a 17m aperture telescope
using state-of-the-art technology has also been proposed. These
projects may achieve thresholds as
low as 20-30 GeV where they will effectively close the current gap
in the $\gamma$-ray spectrum from 20 to 200 GeV. Ultimately this
gap
will be covered by GLAST, the next generation $\gamma$-ray space
telescope (which will use solid-state detectors) which is scheduled
for launch in 2005 by an international collaboration. Extension to
even higher energies can be achieved by the atmospheric Cherenkov
telescopes working at large zenith angles and by particle arrays at
very high mountain altitudes. An interesting telescope that will
soon come on line and will complement these techniques is the
MILAGRO water Cherenkov detector in New Mexico which will operate
24 hours a day with wide field of view and will have good
sensitivity to $\gamma$-ray
bursts and transients.

VERITAS, with seven 10 m telescopes arranged in a hexagonal pattern
with 80 m spacing, will aim for the middle
ground, with its primary objective being high sensitivity in the
100 GeV to 10 TeV range. It will be located in southern Arizona and
will be the logical development of the Whipple telescope. It is
hoped to begin construction in 1999 and to complete the array by
2004.

\begin{figure}
\centerline{\epsfysize 3 truein \epsfbox{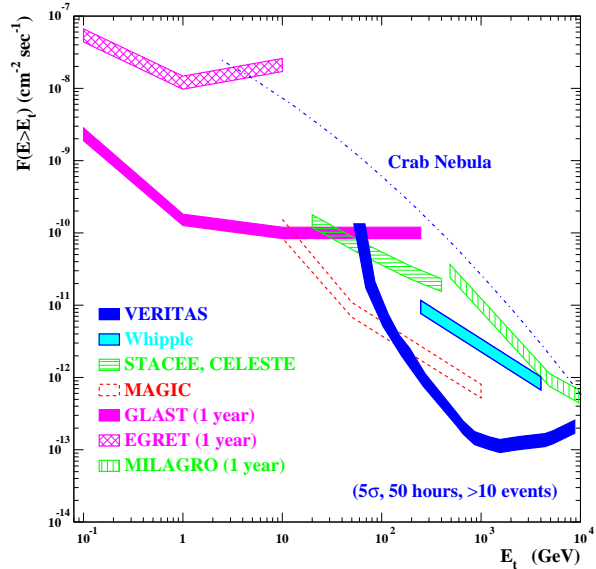}}
\caption{Comparison of the point source sensitivity of VERITAS to
Whipple \protect\cite{Weekes89}, MAGIC \protect\cite{Barrio98},
CELESTE/STACEE \protect\cite{Quebert95}; \protect\cite{Bhat97}; GLAST
\protect\cite{Leonard96}, EGRET \protect\cite{Egret}, and MILAGRO
\protect\cite{Sinnis95}.
\label{sens-fig}
}
\end{figure}

The German-French HESS (initially four, and
eventually perhaps sixteen, 10m class telescopes) will be built in
Namibia and the Japanese NEW CANGAROO array (with three to four
telescopes in Australia) will have similar objectives.  In each
case the arrays will exploit the high sensitivity of the imaging
ACT and the high selectivity of the array approach. The relative
flux sensitivities of the present and next generation of VHE
telescopes as a function of energy are shown in
Figure~\ref{sens-fig}, where the sensitivities of the wide field
detectors are for one year and for the ACT for 50 hours; in all
cases a 5$\sigma$ point source detection is required. The VERITAS
sensitivity is derived from Monte Carlo simulations using the
Whipple telescope as a baseline \cite{Vassiliev98}. The projected
sensitivities of MAGIC, HESS, New CANGAROO and VERITAS are somewhat
similar and we will refer to them collectively as Next Generation
Gamma Ray Telescopes (NGGRTs).

It is apparent from this figure that on the low energy side, the
NGGRTs such as VERITAS will complement the GLAST mission (launch
date 2005) and will overlap with STACEE and CELESTE which
will be coming on line in 1999. At its highest energy, they will
overlap with the Tibet Air Shower Array. It will cover the same
energy range as MILAGRO but with greater flux sensitivity; however
the wide field coverage of MILAGRO will permit the detection of
transient sources which, once detected, can be monitored by
VERITAS. As a Northern Hemisphere telescope VERITAS will complement
the coverage of neutrino sources discovered by AMANDA and ICE CUBE
at the South Pole. Finally if the sources of ultra-high energy
cosmic rays discovered by HiRes and Auger are localized to a few
degrees, VERITAS will be the most powerful instrument for their
further localization and identification.

\subsection{Science to come}
\paragraph{AGNs}
By measuring the high energy end of the spectra for several EGRET
sources. The NGGRTs can help determine what particles produce the
$\gamma$-ray
emission in blazars (electrons should show cut-offs which correlate
with lower energy spectra, protons would not show a simple
correlation).  In addition, the recent efforts \cite{Ghisellini98}
to unify the different
classes of blazar into different manifestations of the same object
type can be tested. In addition the infrared background will be
probed by the detection of sources over a range of redshifts.

\paragraph{SNRs}
The existing data clearly indicate that in order to resolve the
contributions of the various $\gamma$-ray emission mechanisms, one
needs more accurate measurements over a more complete range of
energies.  The NGGRTs and GLAST will be a powerful combination to
address
these issues.  The excellent angular resolution of the NGGRTs will
allow
detailed mapping of the emission in SNRs.  The sensitivity and
energy
resolution, combined with observations at lower
$\gamma$-ray and X-ray energies help to elucidate the $\gamma$-ray
emission mechanism. 
This may lead to direct the confirmation or elimination
of
SNRs as the source of cosmic rays.

\paragraph{Gamma-ray pulsars}

The detection of VHE $\gamma$-rays would be decisive in favoring the
outer gap model over the polar cap model.  Six pulsars are detected
at
EGRET energies and their high energy emission is already seriously
constrained by the VHE upper limits.  The detection of a pulsed
$\gamma$-ray signal above 50 GeV would be a major breakthrough.  

\paragraph{Unidentified galactic EGRET sources}

The legacy of EGRET may be more than 70 unidentified sources, many
of
which are in the Galactic plane.  The positional uncertainty of
these
sources make identifications with sources at longer wavelengths
unlikely.  In the galactic plane, probable sources are SNRs and
pulsars, particularly in regions of high IR density (e.g., OB
associations), but some may be new types of objects. The NGGRT
should
have the sensitivity and low energy threshold necessary to detect
many
of these objects.  Detailed studies
of these objects with the excellent source location capability of
the NGGRTs could lead to many identifications with objects at
longer
wavelengths.  
  Variability
in these objects would be easily identified and measured with
 the NGGRT.

{\bf {Acknowledgements:}}  
Research in VHE $\gamma$-ray astronomy at the Whipple Observatory
is supported by a grant from the U.S.D.O.E. Helpful comments from
Mike Catanese, Stephen Fegan and Vladimir Vassiliev are also
acknowledged.

\end{document}